\documentclass[12pt]{iopart}
\usepackage{times}
\usepackage{graphicx}

\begin{document}

\title{Milne phase for the Coulomb quantum problem related to Riemann's hypothesis}
\author{H C Rosu$^1$, J M Mor\'an-Mirabal$^1$, M Planat$^2$}
\address{Instituto Potosino de Investigaci\'on Cient\'{\i}fica y Tecnol\'ogica, Apdo Postal 3-74 Tangamanga, San Luis Potos\'{\i}, MEXICO}
\address{Laboratoire de Physique et M\'etrologie des Oscillateurs du CNRS, 25044 Besan\c con Cedex, FRANCE}
\begin{abstract}
We use the Milne phase function in the continuum part of the spectrum of the particular Coulomb 
problem that has been employed by Bhaduri, Khare, and Law as an equivalent physical way for calculating the density of zeros of the 
Riemann's function on the critical line. The Milne function seems to be a promising approximate method to calculate the 
density of prime numbers.
\end{abstract}


\noindent
From a 1995 PRE paper of Bhaduri, Khare, and Law [1]  one can obtain the following formula for the density of
zeros of Riemann's zeta function on the critical line 
\begin{equation}
n_{Z}(\epsilon) = -\frac{\ln \pi}{2\pi}+\frac{1}{2\pi}{\rm Re}\Big[\Psi\left(\frac{1}{4}+i\frac{\epsilon}{2}\right)\Big]~,
\end{equation}
where the digamma function is the logderivative of the gamma function, $\Psi(z)=\Gamma ^{'}(z)/\Gamma (z)$.
Formula 16 of the same paper gives the phase shift of a repulsive Coulomb potential obtained
from an inverted oscillator with a hard wall at the origin for the unconventional value of the partial wave number $l=-\frac{1}{4}$. Taking 
the derivative of that formula, one obtains 
\begin{equation}
n_{C}(\epsilon) = -\frac{F^{'}(\epsilon)}{2\pi}+\frac{1}{2\pi}{\rm Re}\Big[\Psi\left(\frac{1}{4}+i\frac{\epsilon}{2}\right)\Big],
\end{equation}
where
\begin{equation}
F(\epsilon)=\frac{\pi}{2}-\tan ^{-1}({\rm cosech} \pi \epsilon).
\end{equation}
Thus, under an appropriate shift, the two expressions given by (1) and (2) differ only by an exponentially small term as noted by the Indian team of
authors.

Our aim in this work is to apply another technique based on the so-called Milne phase function for the calculation of the `density
of states' in the continuum of the same Coulomb problem. Indeed, being a phase, one might think a priori that Milne's function has something to do 
with the nontrivial zeros of the Riemann function. Not only this is a different procedure, but it might be a quite competitive approximation for $n_{Z}$.
Previously, Korsch [2] applied the same method with very good results in the case of bound states for 
a few illustrative cases of elementary quantum mechanics.
He also established that in this approach the density of quantum states is nowhere unique and not even necessarily 
positive (!?). For the Coloumb problem under focus here the 
nonuniqueness issue is actually an advantage because in a certain sense one can choose by trial and error a better 
Milne approximation for $n_{Z}$.
The Milne function is expressed through the following formula
\begin{equation}
n_{M}(y,\epsilon) \equiv \frac{1}{\rho ^2}= \frac{1}{\Big[\alpha \phi  _1(y,\epsilon)+\beta \phi _2(y,\epsilon)\Big]^2+\frac{1}{\alpha ^2}\phi_2^2(y)}~,
\end{equation}
where $\rho$ is the solution of the Pinney nonlinear equation and $\phi _1$ and $\phi _2$ are the two linear independent solutions of the repulsive 
Coulomb problem, i.e.
\begin{equation}
\phi _1(y, \epsilon) =\sin\left(ky-\frac{\epsilon}{2}\ln(2ky)-\frac{l_{r}\pi}{2}+{\rm Arg}\Gamma\left(l_r+1+\frac{i\epsilon}{2}\right)\right),
\end{equation}
and
\begin{equation}
\phi _2(y,\epsilon) =\cos\left(ky-\frac{\epsilon}{2}\ln(2ky)-\frac{l_r\pi}{2}+{\rm Arg}\Gamma\left(l_r+1+\frac{i\epsilon}{2}\right)\right),
\end{equation}
where, as we mentioned before, $l_r=-\frac{1}{4}$. The other symbols are as follows:
$\epsilon=-\frac{E}{\hbar \omega}$ is a reduced spectral parameter, where $E$ is the spectral parameter in the initial inverted oscillator 
problem and $\omega$ is the angular frequency of the oscillator; $k=\frac{m\omega}{2\hbar}$; $y=x^2$, where $x\geq 0$ is the oscillator coordinate.  
The superposition constants $\alpha$ and $\beta$ are determined by arbitrary `initial' conditions at some point $y_0$. We 
fix them in the most `economic' way, which is at the point $y_0=\frac{1}{2k}$ allowing one to eliminate the logarithm in the argument of
the trigonometric functions. Moreover, we employ the Eliezer-Gray prescription, for details see the master thesis of Espinoza [3]. Thus
\begin{equation}
\alpha =\phi _1(\frac{1}{2k}) =\sin\left(\frac{1}{2}+\frac{\pi}{8}+{\rm Arg}\Gamma\left(\frac{3}{4}+\frac{i\epsilon}{2}\right)\right),
\end{equation}
\begin{equation}
\beta =\frac{d\phi _1}{dy}|_{y=\frac{1}{2k}} =(1-\epsilon)\cos\left(\frac{1}{2}+\frac{\pi}{8}+{\rm Arg}\Gamma\left(\frac{3}{4}+\frac{i\epsilon}{2}\right)\right)~.
\end{equation}
A three-dimensional plot of $n_M(y,\epsilon)$ is displayed in Fig.~1, where an expected oscillatory behaviour can be seen.
Not only the procedure based on the Milne phase function could compete very well with other approximate methods for the density of prime numbers but there is a
further advantage on which we briefly comment in the following.  As is well known, the Milne phase function enters as a basic
ingredient in the Ermakov-Lewis phase-amplitude approach for parametric oscillator problems (for recent applications see [4]). To transform the Coulomb problem at hand
\begin{equation}
\frac{d^2\phi}{dy^2}+\Big[k^2-\frac{k\epsilon}{y}+\frac{3}{16y^2}\Big]\phi=0
\end{equation}
into a parametric dynamical problem for a unit mass classical particle, 
one can use the well-known map to canonical classical variables $\phi =q$ and  $\frac{d\phi}{dy}=p$ leading to
\begin{equation}
\dot{q}\equiv \frac{dq}{dy}=p
\end{equation}
\begin{equation}
\dot{p}\equiv\frac{dp}{dy}=-\Big[k^2-\frac{k\epsilon}{y}+\frac{3}{16y^2}\Big]q~,
\end{equation}
where the coordinate $y$ plays the role of the classical Hamiltonian time. Various quantities, such as the Ermakov-Lewis invariant and geometrical angles can be 
calculated easily for this particularly interesting Coulomb problem (due to its connection with prime numbers). But we want to emphasize a different point here. In the
parametric oscillator interpretation and of $y$ as Hamiltonian time, one can use $n_M(y,\epsilon)$ as a direct tool for time-energy
(and therefore time-imaginary axis) characterization of the density of prime numbers as can be inferred from Fig.~1.

\end{document}